# Effect of Silicon atom doping in SiN$_x$ resistive switching films

A. Mavropoulis, N. Vasileiadis, C. Bonafos, P. Normand, V. Ioannou-Sougleridis, G. Ch. Sirakoulis, P. Dimitrakis

*Abstract—* **Doping stoichiometric SiN$_x$ layers (x=[N]/[Si]=1.33) with Si atoms by ultra-low energy ion implantation (ULE-II) and annealing them at different temperatures can significantly impact the switching characteristics. Electrical characterization and dielectric breakdown measurements are used to identify the main switching properties and performance of the fabricated devices and the effect that the Si dopants and annealing temperature have. In addition, impedance spectroscopy measurements revealed the dielectric properties of the silicon nitride films, as well as the ac conductance, which is utilized to identify the conduction mechanisms.**

## I. Introduction

Silicon nitride (Si$_s$N$_4$) is widely used for charge-trapping (CT) nonvolatile memories for more than half of century [1, 2] in Flash SONOS [3] and Vertical CT-NVM technologies [4]. The thermodynamics of the deposition conditions result in a nitrogen deficiency in the amorphous SiN$_x$ layer that causes intrinsic defects of different configurations to be formed, i.e., nitrogen vacancies, Si-dangling bonds etc. [5]. Recently, the CT-NVM technology was utilized to implement neuronal synapses. Furthermore, functional resistive memory (RRAM) cells with SiN$_x$ as active material have been demonstrated [6, 7]. The effect of nitride layer's stoichiometry x=[N]/[Si] (x=1.33 for Si$_3$N$_4$) as well as the nature of the electrode material have been investigated in [6, 7] and [7, 8] respectively. The very attractive properties of silicon nitride render it a great solution for memory cell scaling [9], neuromorphic, in-memory and edge computing [10, 11, 12] as well as security applications [13, 14] realizing true-random number generators and physical unclonable functions. It is possible to program SiN$_x$-memristors at different low-resistance states (LRS) well distinguished with respect to the high-resistance state (HRS) resulting in a large operation window.

The resistance switching characteristics of the RRAMs can be controlled by doping the dielectric materials, a behavior that has been investigated [15, 16].

The technological requirements are continually escalating, necessitating devices with varying characteristics to meet the demands of different applications [17]. In this work, ultra-low energy ion implantation (ULE-II) of Si atoms is performed on almost stoichiometric SiN$_x$ layers which are then annealed using temperatures in inert ambient. The effect of Si doping in the resistance switching properties and memory performance is investigated by utilizing thorough material and memory cell characterization.

## II. Experimental

Firstly, 6.0nm ± 0.5nm LPCVD SiN$_x$ (x=1.27) layers were deposited on three heavily Phosphorus-doped Si wafers (N$_d$=5×10$^{19}$ P/cm$^{-3}$) [16]. After that each wafer was implanted at 3keV with different doses of Si-ions. To heal the ULE-II damage and incorporate Si atoms in the SiN$_x$ lattice a low temperature (low-T) and a high temperature (high-T) post-implantation annealing (PIA) were performed, as shown in Table 1. With dissimilar processing steps the formation of Si nanocrystals was achieved in amorphous SiN$_x$ layers by ULE-II [16]. Specifically, the temperatures of 800°C and 950°C were chosen based on preliminary studies indicating optimal healing of implantation damage and effective incorporation of Si atoms into the SiN$_x$ lattice. Further increase in the annealing temperature was not pursued in this study but will be considered for future research to explore the potential effects. Fig. 1 depicts the structure of the examined samples and the calculated TRIM profiles of the implanted Si-atoms are shown in the inset of Fig.1. The top-electrode (TE) is created from a 30nm Cu layer covered by 30nm Pt layer to protect from oxidation. HP4155 and Tektronix 4200A were used for dc electrical characterization of the RRAM cells and impedance measurements were carried out using HP4284 and Zurich Instruments MFIA. For all the measurements the bottom electrode was grounded, and the voltage was applied on the top electrode. A wafer prober was used to make the connections to the samples, which were measured at room temperature.

* This work was financially supported by the research project "LIMA-chip" (Proj.No. 2748) of the Hellenic Foundation of Research and Innovation (HFRI).

A. Mavropoulis, N. Vasileiadis, P. Normand, V. Ioannou-Sougleridis and P. Dimitrakis are with the Institute of Nanoscience and Nanotechnology, NCSR "Demokritos", Ag. Paraskevi 15341, Greece

N. Vasileiadis and G. Ch. Sirakoulis are with the Department of Electrical and Computer Engineering, Democritus University of Thrace, Xanthi 67100, Greece

C. Bonafos is with the CEMES-CNRS et Université de Toulouse, BP94347-31055 Toulouse, Cedex 4, France

**Table 1** Table of fabricated and examined samples

| Sample Id. | S12 | S13 | S22 | S23 | S42 | S43 |
|---|---|---|---|---|---|---|
| I.I. Dose ($Si^+/cm^{-3}$) | $1\times10^{14}$ | | $5\times10^{14}$ | | $5\times10^{15}$ | |
| Furnace Annealing | 800°C 20min $N_2$ | 950°C 20min $N_2$ | 800°C 20min $N_2$ | 950°C 20min $N_2$ | 800°C 20min $N_2$ | 950°C 20min $N_2$ |

### III. RESULTS AND DISCUSSION

To start with, I-V sweeps were carried out using different current compliances during the SET, $I_{CC}$, to study the SET/RESET characteristics. The I-V curves for all the samples, as well as the stoichiometric for comparison, with $I_{CC} = 100\mu A$ are presented in Fig. 1. There is a significant change in the current density of the Si-doped devices before switching. More specifically, the current before SET and after RESET is reduced for the $SiN_x$ implanted with the lowest and highest dose. This leads to an increase in the SET ($V_{SET}$) and RESET ($V_{RESET}$) voltages. This behavior is attributed to the thickness increase after the doping process due to the incorporation of extra Silicon atoms and will be discussed later.

After statistical analysis of the $V_{SET}$ and $V_{RESET}$ voltages, the results were compared by calculating the coefficient of variation σ/μ, which is defined as the ratio of the standard deviation σ over the mean value μ. This ratio is almost the same for all samples for the $V_{SET}$. On the other hand, for the $V_{RESET}$ this ratio is lower for samples annealed at a high-T compared to samples annealed at low-T. We can conclude that annealing at 950°C is more effective at healing the implantation damage and it allows the implanted Si-atoms to be integrated more effectively with the lattice atoms. This causes the reduction of the cell-to-cell variation and consequently the reduction of the $V_{SET}$ and $V_{RESET}$ variation after each operation cycle. Moreover, in Fig. 1 it becomes clear that by increasing the implantation dose the LRS decreases. This means that the increase in the excess Si in the material leads to the easier formation of the conductive filament. It is interesting to note that the mechanism responsible for the resistance switching is related to the traps in the material and more specifically to their concentration. These traps in the $SiN_x$ films mainly include the various forms of nitrogen vacancies and the resulting Si dangling bonds which under the effect of an electric field create conductive filaments that short the two electrodes and allow electrons to flow through them [18].

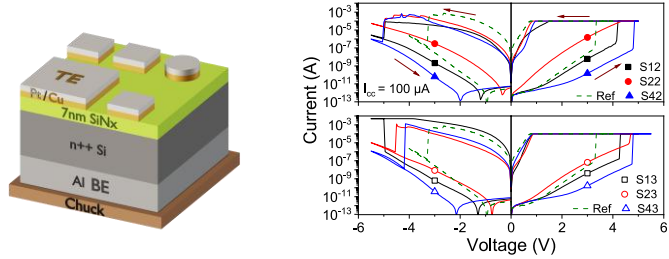

**Fig. 1.** Device structure description and comparative plots of I-V sweeps

In the past, the conduction mechanisms in $SiN_x$ films for memristive devices has been investigated and most research investigations conclude that the mechanism is either trap-limited (space charge limited conduction, SCLC) [8, 15, 19] or bulk-limited (Poole-Frenkel conduction, P-F) [9] or even a combination of these mechanisms (modified SCLC, MSCLC) [20]. The following equations are used to describe the previously mentioned conduction mechanisms:

$$I = \frac{9}{8}A\mu\varepsilon\varepsilon_0\theta\frac{V^2}{d^3} \quad \text{where} \quad \theta = \frac{N_C}{N_T}e^{-q\varphi_T/k_BT} \qquad \text{for SCLC (1)}$$

$$I = q\mu N_C A\left(\frac{V}{d}\right)exp\left[\frac{q(\varphi_T - \sqrt{qV/d\pi\varepsilon\varepsilon_0})}{k_BT}\right] \qquad \text{for P-F (2)}$$

$$I = \alpha V^2 e^{\beta\sqrt{V}} \qquad \text{for MSCLC (3)}$$

$$\alpha = \frac{9\varepsilon\varepsilon_0\mu N_C A}{8d^3 N_T}e^{-q\varphi_T/k_BT} \qquad (3a)$$

$$\beta = \frac{q}{k_BT}\sqrt{\frac{q}{\pi\varepsilon\varepsilon_0 d}} \qquad (3b)$$

where A, d, ε, $N_C$, μ are the device area, film thickness, dielectric constant, density of states in the conduction band of the dielectric and the electronic drift mobility respectively and all the other symbols have their usual meaning. Also, $\varphi_T$ and $N_T$ are the trap energy and concentration respectively.

The I-V curves in Fig. 1 at HRS, before SET, are analyzed following the P-F and MSCLC conduction mechanisms and by performing linear fits, it is obvious that both mechanisms fit the data extremely well. However, slightly better fitting coefficients can be achieved with P-F. For the reference non-implanted $SiN_x$ the conduction mechanism is SCLC and has already been published [15]. This significant difference in the conduction mechanisms is caused by the redistribution of traps in the $SiN_x$ after the ULE-II and PIA.

**Table 2** P-F linear fitting slopes and calulated dielectric constants

| Sample | β (V$^{-1/2}$) | ε | Corrected ε |
|---|---|---|---|
| S12 | 11.57 ± 0.05 | 10.07 ± 0.08 | 5.04 ± 0.04 |
| S22 | 10.93 ± 0.02 | 11.29 ± 0.04 | 5.65 ± 0.02 |
| S42 | 12.45 ± 0.06 | 8.70 ± 0.08 | 4.35 ± 0.04 |
| S13 | 10.14 ± 0.02 | 13.12 ± 0.05 | 6.56 ± 0.03 |
| S23 | 10.42 ± 0.02 | 12.42 ± 0.04 | 6.21 ± 0.02 |
| S43 | 11.40 ± 0.03 | 10.38 ± 0.05 | 5.19 ± 0.03 |

P-F plots linearize equation (2) allowing for the calculation of the pre-exponential factor C from the intercept and the dielectric constant ε from the slope β described by equation (3b):

$$C = \frac{q\mu N_C A}{d} exp\left[\frac{q\varphi_T}{k_B T}\right] \quad (4)$$

$$\varepsilon = \frac{q^3}{\pi\varepsilon_0 d} \frac{1}{(\beta k_B T)^2} \quad (5)$$

To calculate the trap energy and electronic drift mobility in silicon nitride films, measurements at various temperatures are required. It should be noted that C increases with trap density: the more silicon-rich the $SiN_x$ films become, the more the conductivity and the value of C increases. However, the electronic drift mobility is responsible for the significant difference observed in C while the nitride becomes more silicon-rich [21]. The increase in C is followed by an increase in the current density for an increasing ε. The calculated slopes of the P-F plots, as well as the calculated dielectric constants are presented in Table 2. The extracted ε values differ significantly from those in the literature for $SiN_x$, which are typically in the range of 5 – 7.5.

The fabricated samples are investigated even further by utilizing impedance spectroscopy measurements using a small ac signal of 25mV at SET (LRS) and RESET (HRS) states for all devices listed in Table 1, as well as a +0.1V dc bias. The impedance was measured for samples at the LRS after a SET with $I_{CC}$ = 100μA. The analysis of the impedance revealed the dependence of the dielectric constant ε'=Re(ε*)=ε and ac conductance σ'=Re(σ*) for the investigated nitride films.

**Table 3** LRS impendance spectroscopy results ($I_{CC}$ = 100μA)

| Sample | S12 | S22 | S42 | S13 | S23 | S43 | Ref. |
|---|---|---|---|---|---|---|---|
| $R_p$ (MΩ) | 1.02 | 0.39 | 0.51 | 0.54 | 0.83 | 0.59 | 0.008 |
| $C_p$ (pF) | 60.4 | 57.6 | 62.5 | 57.3 | 57.9 | 55.0 | 72.5 |
| ε' @ 1kHz | 4.95 | 4.66 | 5.07 | 4.57 | 4.62 | 4.42 | 5.82 |

The values of ε from the impedance spectroscopy measurements in Table 3 are the correct ones and appear to be significantly lower compared to the these calculated from the P-F slopes in Table 2. This large difference is created from the lack of compensation during the calculation of donor-traps by acceptor-traps [21], that is known also as "anomalous P-F effect" [22]. The existence of acceptor sites results in a change in the slope of the ln(I/V) vs. $V^{1/2}$ plot. This is attributed to the change of Fermi level with respect to the defect/trapping levels, suggesting that both C and ε are affected. This change can be accommodated by using an acceptor compensation factor [21, 22]. So, the original equation (2) is modified as follows:

$$I = q\mu N_C A \left(\frac{V}{d}\right) exp\left[\frac{q(\varphi_T - \sqrt{qV/d\pi\varepsilon\varepsilon_0})}{\xi k_B T}\right] \quad (6)$$

The compensation factor ξ is added in the denominator of the exponential, where ξ=2 when no acceptor traps exist and ξ=1 when there is a considerable number of acceptor traps. In general, ξ, can range between 1 and 2 depending on the position of the Fermi level [23, 24]. Therefore, the values of ε in Table 2 are modified and resemble more closely the values obtained by impedance spectroscopy. The small variations observed are due to use of the same film thickness d during calculations. However, it has been proven that the thickness is increased with the incorporation of Si dopants by ULE-II (swelling effect) [16]. Indeed, XTEM measurements revealed the accurate value of d for high-T annealed samples: 5.6, 6.3 and 7.3 nm for S13, S23 and S43 samples respectively.

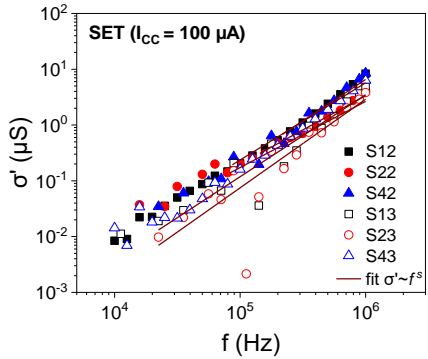

**Fig. 2.** Frequency dependence of AC conductivity for all examined samples.

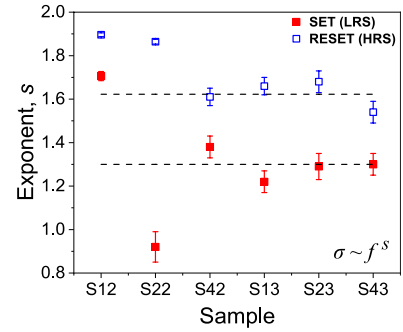

**Fig. 3.** AC conductivity exponent *s* for different samples

The ac conductance is also calculated. The experimental results include the dc conductance and by subtracting the dc part, the ac conductance can be extracted and is shown in Fig.2. We can fit these results and it becomes evident that for high-T annealed layers σ' varies as $\sim f^s$, where s is close to 1.3 and 1.62 for LRS and HRS respectively (Fig. 3). According to [23], the above values of s denote that during SET and RESET the conduction in doped-SiN$_x$ layers is mainly governed by electron variable range hopping (s is close to 1) and trap-to-trap tunneling mechanisms (s is close to 2) respectively. Moreover, a slight deviation from the linear trend is observed in lower frequencies. This is where the corner frequency is located, beyond which the conductance increases [23], due to the inability of the dipoles to align with the ac signal.

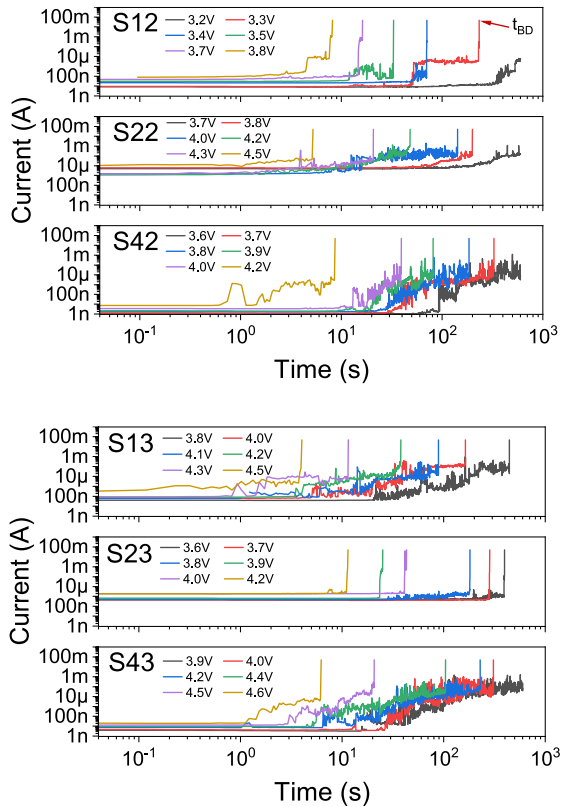

**Fig. 4.** Time to breakdown measurements employing constant voltage stress experiments.

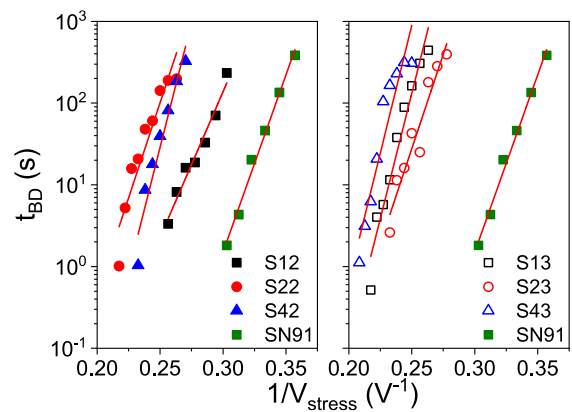

**Fig. 5.** Time of breakdown vs $1/V_{Stress}$ for different samples.

Finally, time dependent dielectric breakdown tests on pristine devices were performed. In these experiments, a constant voltage was applied on a device, which is in the HRS, and the current was measured every 100ms. The current evolution through time at various constant voltages is shown in Fig. 4. From these measurements, time-to-breakdown ($t_{BD}$) was measured. In Fig. 5, the $t_{BD}$ dependence on the $1/V_{Stress}$ is demonstrated. According to [24]

$$t_{BD} = Ae^{B(d-\Delta d)/V_{Stress}} \qquad (7)$$

where $E_{ox}$ (= $V_{Stress}$/d) is the electric field, $\Delta d$ is the effective dielectric thinning due to the defects formed inside the $SiN_x$ and B is a fitting parameter. The slope of linear least-square fitting in Fig. 5 is called the electrical field acceleration factor and is equal to

$$B(d-\Delta d)/V_{Stress} \qquad (8)$$

This slope is almost the same for the high-T annealed samples. For the low-T annealed samples, it increases with the ion implantation dose. In general, the dielectric strength (the voltage required for breakdown) is higher for high-T annealed devices. This is attributed to the reduction of the defects in the material that serve as flaws which cause dielectric breakdown. It is also interesting to note that stoichiometric $SiN_x$ is thinner than the doped layers and breaks down at lower voltages, which is in line with the lower $V_{SET}/V_{RESET}$ mentioned before.

## IV. Conclusions

The switching characteristics of ULE-II Si-doped $SiN_x$ thin layers were investigated. The conduction process is mainly governed by donor-like traps caused from the large density of trivalent Si-dangling bonds or similar defects. AC conductance measurements allow the identification of the conduction mechanism at LRS and HRS respectively. Moreover, annealing at high-T improves the breakdown characteristics of the material.

## Declaration of Competing Interest

The authors declare the following financial interests/personal relationships which may be considered as potential competing interests: This work was financially supported by the research project "LIMA-chip" (Proj.No. 2748) of the Hellenic Foundation of Research and Innovation (HFRI).